\documentclass[aip,reprint,rsi]{revtex4-1}
\usepackage[utf8]{inputenc}
\usepackage{graphicx}
\usepackage[load-configurations=abbreviations,detect-mode,separate-uncertainty]{siunitx}
\usepackage[colorlinks=true,linkcolor=blue,citecolor=blue,urlcolor=blue]{hyperref}
\usepackage{amsmath}
\usepackage{amssymb}

\begin{document}

\title{Ultrafast supercontinuum fiber-laser based pump-probe scanning MOKE microscope for the investigation of electron spin dynamics in semiconductors at cryogenic temperatures with picosecond time and micrometer spatial resolution}

\author{T.~Henn}
\author{T.~Kießling}
\email[E-mail: ]{tobias.kiessling@physik.uni-wuerzburg.de}
\author{W.~Ossau}
\author{L.~W.~Molenkamp}
\affiliation{Physikalisches Institut (EP3), Universität Würzburg, 97074 Würzburg, Germany}
\author{K.~Biermann}
\author{P.~V.~Santos}
\affiliation{Paul-Drude-Institut für Festkörperelektronik, Hausvogteiplatz 5-7, 10117 Berlin, Germany}

\date{\today}

\begin{abstract}
We describe a two-color pump-probe scanning magneto-optical Kerr effect (MOKE) microscope which we have developed to investigate electron spin phenomena in semiconductors at cryogenic temperatures with picosecond time and micrometer spatial resolution. The key innovation of our microscope is the usage of an ultrafast `white light' supercontinuum fiber-laser source which provides access to the whole visible and near-infrared spectral range. Our Kerr microscope allows for the independent selection of the excitation and detection energy while avoiding the necessity to synchronize the pulse trains of two separate picosecond laser systems. The ability to independently tune the pump and probe wavelength enables the investigation of the influence of excitation energy on the optically induced electron spin dynamics in semiconductors. We demonstrate picosecond real-space imaging of the diffusive expansion of optically excited electron spin packets in a (110) GaAs quantum well sample to illustrate the capabilities of the instrument. 

\end{abstract}


\maketitle
\section{Introduction}
The investigation of electron spin phenomena has developed into an important field of solid state physics in the recent past. Spin-sensitive magneto-optical spectroscopy has emerged as a particularly successful technique for the exploration of the electron spin dynamics in bulk and low-dimensional semiconductors. Taking advantage of the spin selectivity of the optical interband transitions in zincblende-type semiconductors\cite{OpticalOrientation}, spin-polarized electrons are excited by above-barrier illumination of the sample with a circularly polarized pump laser. In its simplest form, magneto-optical spectroscopy infers the state of the electron spin system by using the degree of circular polarization of the photoluminescence as a tracer for the electron spin polarization. However, photoluminescence spectroscopy depends on the simultaneous presence of both electrons and holes and is therefore limited to time and length scales of the order of the minority carrier lifetime and the ambipolar charge diffusion length.

A more sophisticated approach, pump-probe magneto-optical Kerr effect (MOKE) spectroscopy, eludes this limitation by deducing the electron spin polarization from a measurement of the change of the major axis of optical polarization, the Kerr rotation, of a second probe laser which is reflected from the sample surface. Using fast pulsed laser sources and strong focussing optics, pump-probe MOKE microscopy allows for the investigation of the electron spin system with sub-picosecond time and micrometer spatial resolution with ultra-high sensitivity.

Pump-probe MOKE spectroscopy has been used extensively to study electron and hole spin transport and relaxation processes in bulk semiconductors, quantum wells, and quantum dots. Notable applications of magneto-optical Kerr and Faraday microscopy include the observation of electron spin lifetimes of the order of \SI{100}{\ns} (Ref. \onlinecite{Kikkawa:1998}) and spin drift over distances exceeding \SI{100}{\micro\meter} (Ref. \onlinecite{Kikkawa:1999}) in bulk n-GaAs, the study of the influence of electric, magnetic, and strain fields on optically excited electron spins,\cite{Crooker:2005} the optical investigation of electrical spin injection in n-GaAs channels,\cite{Crooker:2005tx} the observation of the spin Hall effect in bulk\cite{Kato:2004} and low-dimensional semiconductors,\cite{Sih:2005} and the magneto-optical detection of single electron spins in quantum dots.\cite{Berezovsky:2006}

In the traditional form of degenerate pump-probe spectroscopy, the pulsed pump and probe lasers are obtained from the same laser source, i.e. the excitation and detection wavelengths are equal. In semiconductors, an appreciable Kerr rotation signal is only observed in the spectral vicinity of the excitonic optical resonances. The necessity to detect the small polarization signals therefore limits the excitation energies accessible by degenerate MOKE spectroscopy. However, a strong influence of the excitation energy on the electron spin dynamics due to pump-induced carrier heating and the resulting electron temperature gradients has been observed in the past both for bulk semiconductors\cite{Quast:2013,Henn:2013,Henn:2013tr} and quantum wells.\cite{Weber:2005,Carter:2006} It is therefore desirable to independently control the pump and probe wavelengths in MOKE spectroscopy experiments to enable the systematic investigation of the interrelation between charge, spin and heat transport in the electron system of semiconductor heterostructures.

In the past this ability has been achieved by elaborate synchronization of the pulse trains of two independent Ti:sapphire laser systems.\cite{Crooker:1996,Kikkawa:1997} Here we present a new approach to time-resolved two-color pump-probe MOKE microscopy which is based on an ultrafast `white light' supercontinuum fiber-laser source. Such fiber-based supercontinuum sources have become commercially available in the recent past and are increasingly used in optical spectroscopy experiments. 

The three main advantages of our instrument are (i) the ability to independently tune the excitation and detection energy while avoiding the necessity to synchronize two separate picosecond laser systems, (ii) access to the whole visible and near-infrared spectral range for the investigation of a wide range of different semiconductor material systems, and (iii) a significant reduction of the cost involved in the development of a picosecond two-color Kerr microscope due to the utilization of a single, comparatively inexpensive femtosecond fiber-laser source.

We have developed the Kerr microscope to study low-temperature electron spin transport by means of time-resolved real-space imaging of the diffusion of optically excited electron spin packets in bulk semiconductors and quantum wells (QWs). Typical spin diffusion lengths in semiconductors are of the order of several micrometers.\cite{Crooker:2005,Furis:2007} In bulk n-GaAs long spin relaxation times exceeding \SI{100}{\ns} are routinely observed.\cite{Kikkawa:1998,Dzhioev:2002,Romer:2010} However, spin relaxation in QWs is  much faster and typically happens on time scales of the order of hundred picoseconds.\cite{Damen:1991,Ohno:1999,Korn:2010} Time-resolved real-space imaging of low-temperature electron spin diffusion processes therefore requires an instrument which allows to investigate electron spins with micrometer spatial and picosecond time resolution at cryogenic temperatures. The pump-probe MOKE microscope presented here meets these requirements and is capable of measuring optically induced electron spin diffusion for sample temperatures between \SI{8}{\kelvin}-\SI{300}{\kelvin} with $\approx\SI{1}{\ps}$ time and $\lesssim\SI{2}{\micro\meter}$ spatial resolution.

To illustrate the capabilities of the instrument, we demonstrate picosecond real-space imaging of the diffusive expansion of an optically excited electron spin packet in a (110) GaAs quantum well sample. From the time-resolved measurement of the increase of the spin packet width we directly determine the spin diffusion coefficient $D_{s}$ as a function of excitation power and lattice temperature.

\section{Kerr microscopy setup}
\begin{figure*}[htb]
\includegraphics[width=1.5\columnwidth]{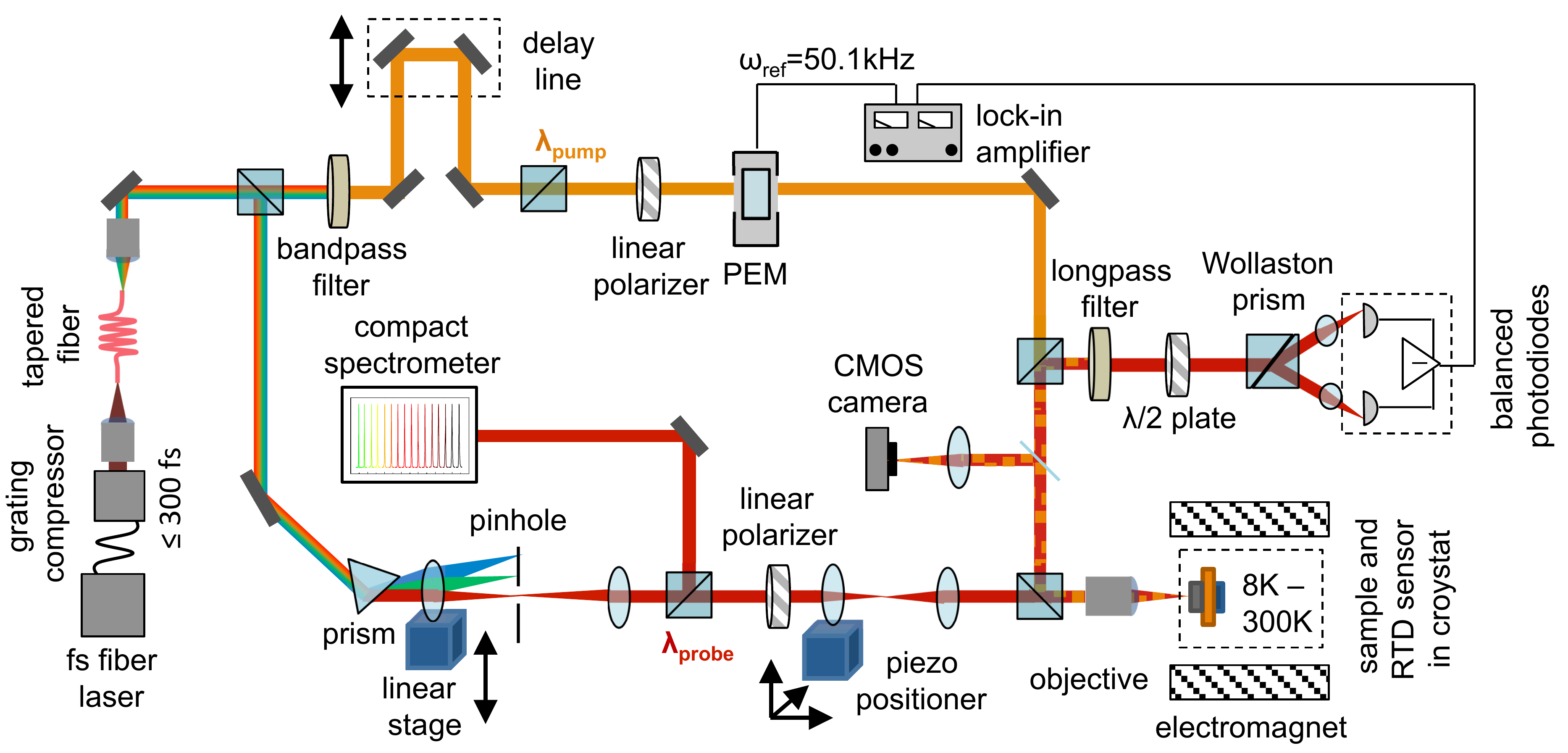}
\caption{Schematic of the two-color pump-probe MOKE microscope. The main optical components of the Kerr microscope are the supercontinuum generation, the instrumentation for the independent spectral filtering of the pump and probe laser, the focussing and raster-scanning optics for the local optical electron spin excitation and detection and the manipulation of the probe laser position, and the balanced photoreceiver lock-in Kerr rotation detection scheme for the measurement of the small spin-dependent changes of the probe laser polarization.}
\label{fig1}
\end{figure*}
In the following we provide a detailed description of the two-color pump-probe supercontinuum MOKE microscope. A diagram of the experimental setup is shown in Fig. \ref{fig1}. To minimize disturbances by dynamic changes in the ambient environment we operate the Kerr microscope in a climate-controlled laboratory where we stabilize the temperature and relative humidity within $\pm\SI{1}{\celsius}$ and below $\SI{35}{\percent}$, respectively. The instrument is built on an optical table equipped with an active vibration isolation system. The detrimental effects of turbulent air currents on the polarization state of the pump and probe lasers are reduced by enclosing the entire setup with a laminar flow box.

Access to sample temperatures between \SI{8}{\kelvin} and \SI{300}{\kelvin} is provided by an Oxford MicrostatHe model narrow tail liquid helium flow optical cryostat which enables the utilization of short focal length focussing optics. The cryostat is mounted on a 3-axis stage to allow for lateral scanning of the sample position and focussing of the pump and probe beam.

The sample is yieldably clamped to a copper holder which is mounted on the coldfinger of the cryostat. Care is taken to avoid a strain induced $\mathbf{k}$-dependent splitting of the conduction band spin states which affects the electron spin dynamics.\cite{Crooker:2005} A calibrated Cernox resonant tunneling diode (RTD) temperature sensor is attached to the backside of the copper holder for accurate measurements of the sample temperature. The cryostat is placed between the poles of an electromagnet which allows for the application of external in-plane magnetic fields of up to \SI{800}{\milli\tesla} in Voigt geometry. A calibrated Hall sensor is used to monitor the external magnetic field at the sample position.

The main optical components of the pump-probe MOKE microscope are the supercontinuum generation, the instrumentation for the independent spectral filtering of the pump and probe laser, the focussing and raster-scanning optics for the local optical electron spin excitation and detection and the manipulation of the probe laser position, and the balanced photoreceiver lock-in detection scheme for the measurement of the small spin-induced changes of the probe laser polarization. A detailed description of each component is given in the following.

\subsection{Supercontinuum generation}
The generation of the `white light' supercontinuum from which we derive the pump and probe laser beams is based on nonlinear frequency conversion of a short near-infrared laser pulse in a tapered photonic fiber. For the optical pumping of the fiber we use the output of a mode-locked PolarOnyx Uranus 1030 high power femtosecond laser which provides pulses with a central wavelength of $\approx\SI{1040}{\nm}$. The repetition rate of the pump laser system is \SI{36.4}{\mega\hertz} (\SI{27.5}{\ns} pulse-to-pulse interval). The time-averaged output power of the pump laser is \SI{3}{\watt}. After exciting the pump laser, a grating compressor is used to reduce the output pulse length to $\lesssim\SI{300}{\femto\second}$.

The pump pulse is coupled into the tapered fiber by an infinity corrected Olympus 10x plan achromat objective (numerical aperture $\text{NA}=\SI{0.25}{}$) which is mounted on a 3-axis fiber launch system. A Faraday isolator is placed in the pump beam path in front of the objective to prevent unintentional back-reflection of the pump light into the laser cavity. Owing to the high peak intensities achieved by the strong femtosecond pump laser in the thin tapered fiber, a variety of non-linear effects including soliton fission, stimulated Raman and Brillouin scattering, four-wave-mixing, and self-phase modulation lead to a strong spectral broadening of the pump pulse.\cite{Dudley:2006} Depending on specific parameters such as fiber waist diameter, excitation pulse duration, pump wavelength, and peak pulse power this spectral broadening results in the creation of a smooth supercontinuum which can extend from $\approx\SI{400}{\nm}-\SI{1600}{\nm}$.\cite{Teipel:2003,Teipel:2005} 

The configuration of the instrument described here is optimized for the investigation of GaAs-based semiconductor heterostructures. We therefore use a comparatively thick fiber with a \SI{4}{\micro\meter} waist diameter which leads to supercontinuum creation with appreciable spectral weight in the vicinity of the relevant $\approx\SI{800}{\nm}$ spectral range. The maximum time-averaged power transmitted through the fiber is \SI{880}{\milli\watt}. A typical spectrum of the output of the supercontinuum source is shown in Fig. \ref{fig2} (a) on a logarithmic scale. While the supercontinuum in the present configuration is limited to wavelengths $\gtrsim\SI{550}{\nm}$, access to higher photon energies can be obtained by using a thinner fiber.

After exiting the fiber the supercontinuum light is collimated by a second, infinity corrected, fiber launch mounted Olympus 4x plan achromat microscope objective ($\text{NA}=\SI{0.1}{}$). We use a broadband anti-reflection coated beamsplitter to divide the supercontinuum beam in two components. We obtain the pump and probe lasers from the two beams by performing individual spectral filtering.

\subsection{Spectral filtering of the pump and probe beam}
The spectral position of the excitonic Kerr resonance depends on the semiconductor material, the sample temperature, and in the case of low-dimensional systems on the confinement energy. While it is usually sufficient to operate the pump laser at fixed above-bandgap wavelengths, MOKE microscopy requires a continuously tunable probe laser to allow for a versatile investigation of different semiconductor heterostructures.

To meet this requirement we have implemented a prism-based spectral filtering scheme for the probe laser. The probe beam is horizontally dispersed by a N-SF14 glass prism. A high-dispersion glass type is used to enhance the angular spread of the individual spectral components of the supercontinuum to achieve a high spectral resolution. To minimize reflection losses, the prism is operated under the condition of minimum deviation. The \SI{59.6}{\degree} apex angle of the prism is designed such that the incident and exit angles are made Brewster's angle for wavelengths between $\approx\SI{700}{}-\SI{900}{\nm}$. The lossless transmitted horizontal polarization component of the initially unpolarized supercontinuum is subsequently used as the linearly polarized probe laser beam as described below.

The dispersed probe beam is focused by a $f=\SI{100}{\mm}$ lens on a \SI{50}{\micro\meter} pinhole. In the focal plane the angular dispersion translates to a lateral displacement of the focus position of different wavelength components of the supercontinuum. The focussing lens is mounted on a PI miCos LS-65 linear stage which allows for horizontal scanning of the lens position with sub-micrometer resolution and uni-directional repeatability. Variation of the lens position allows for a selection of the desired wavelength component from the supercontinuum by the pinhole.  After passing the pinhole, the transmitted spectral component is collimated by a second $f=\SI{60}{\mm}$ lens. By manipulating the position of the first lens while keeping the pinhole position fixed, the probe wavelength can be tuned without changing the beam direction after the second lens.

\begin{figure}[tb]
\includegraphics[width=1\columnwidth]{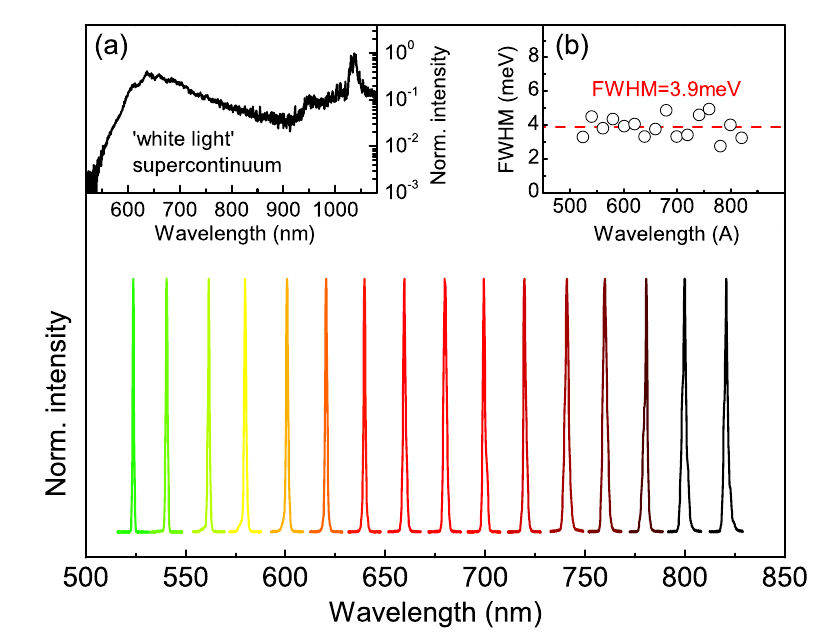}
\caption{Spectra of the continuously tunable probe laser obtained by dispersive filtering of the white-light supercontinuum. Inset (a): Typical output spectrum of the pulsed fiber-laser source (\SI{4}{\micro\meter} fiber waist diameter, \SI{880}{\milli\watt} time-averaged total output power). Inset (b): Spectral width (FWHM) of the probe laser as a function of central laser wavelength $\lambda_{\text{probe}}$. The red dashed line indicates the mean FWHM of \SI{3.9}{\milli\electronvolt}.}
\label{fig2}
\end{figure}

In Fig. \ref{fig2} we exemplarily demonstrate the continuous tuning of the probe laser over a wide spectral range between \SI{520}{\nm} and \SI{820}{\nm} by a systematic variation of the scanning lens position. For each lens position we measure the probe laser spectrum with a compact CCD spectrometer which can be introduced in the probe beam path after the filtering instrumentation. 

We determine the energy resolution of the prism filter by measuring the full-width at half maximum (FWHM) of the probe laser spectrum as a function of the central wavelength $\lambda_{\text{probe}}$. In Fig. \ref{fig2} (b) we show that no systematic variation of the probe laser linewidth with increasing $\lambda_{\text{probe}}$ is observed. The FWHM stays approximately constant over the whole examined wavelength range. The mean value of the FWHM is \SI{3.9}{\milli\electronvolt} and the FWHM never exceeds \SI{5}{\milli\electronvolt}. If necessary, the spectral resolution of the prism filter can be improved by using a smaller pinhole and a higher dispersion glass type;  however this leads to a reduction of the available probe laser power.

The pump laser wavelength is selected by passing the supercontinuum through a standard optical bandpass filter. A broad selection of high-quality dielectric bandpass filters which cover the whole visible and near-infrared spectral range is commercially available. Depending on the material system under investigation, filters with appropriate central wavelengths and passband widths can be introduced in the pump laser beam path. For the measurements presented in this manuscript we employ a dielectric bandpass with a central wavelength $\lambda_{\text{pump}}=\SI{780}{\nm}$ and a spectral FWHM of \SI{10}{\nm} ($\approx\SI{20}{\milli\electronvolt}$).

A synchronized pair of conventional mechanical delay lines is placed in the pump beam path to introduce a variable time delay $\Delta t$ between the pump and probe pulses. The PI miCos LS-110 linear stages each have a travel range of \SI{305}{\milli\meter} with \SI{50}{\nano\meter} resolution and uni-directional repeatability. The present configuration provides a maximum delay of \SI{4}{\ns} which can be extended by multi-pass operation of the delay lines.

\subsection{Focussing and raster-scanning optics}
The diffraction limit for the spatial resolution $\Delta$ of conventional far-field optical microscopy is determined by the wavelength $\lambda$ and the NA of the focussing objective:\cite{BornWolf:2006}
\begin{equation}
\label{ResolutionLimitEq}
\Delta\geq\frac{1}{2}\frac{1.22\lambda}{\text{NA}}
\end{equation}
Achieving micrometer spatial resolution in the near-infrared and visible range therefore requires the use of high NA focussing optics with short focal distances. To minimize aberrations in the spatially resolved electron spin detection it is further desirable to utilize high-quality room-temperature microscope objectives which must be operated outside the cryostat. Spatial constraints dictate to use short working distance optical cryostats and the utilization of the same microscope objective for the focusing of both the pump and probe laser beam.

We use an infinity-corrected Mitutoyo 50x plan apochromatic long working-distance microscope objective ($f=\SI{4}{\mm}$, $\text{NA}=\SI{0.42}{}$) to focus the pump and probe beam at normal incidence on the sample surface. To scan the probe with respect to the fixed pump beam position, we introduce a pair of $f=\SI{4}{\mm}$ aspheric lenses arranged in confocal geometry in the probe beam path. The first lens is mounted on a PI P-611.3 Nanocube 3-axis piezo positioner which offers \SI{100}{\micro\meter} travel range with nanometer resolution along each axis. 

Manipulation of the relative position of the aspheric lens pair changes the angle of incidence on the microscope objective which translates to a change of the focus position of the probe beam on the sample surface. Since the focal lengths of the microscope objective and the aspheric lenses are equal, lateral scanning of the piezo-mounted lens directly results in a translation of the probe laser beam position at the sample surface by the same distance. The Kerr microscope therefore offers a $\SI{100}{}\times\SI{100}{\micro\meter\squared}$ field of view.

After reflection from the sample surface the pump and probe lasers are collected by the same microscope objective. Polarization retaining beamsplitters are used to steer the pump and probe beams to the Kerr rotation detection optics and to a CMOS camera which is placed in the focal plane of a $f=\SI{75}{\milli\meter}$ lens. The CMOS camera is used to monitor the focussing of both lasers and the positioning of the probe beam. We additionally use the CMOS camera for the determination of the spatial resolution of the MOKE microscope as described below.

\subsection{Optical spin excitation}
We employ the standard optical orientation technique\cite{OpticalOrientation} for the electron spin excitation by the pump laser. We use a Hinds Instruments PEM-100 photoelastic modulator\cite{Kemp:1969} (PEM) to periodically modulate the pump polarization between $\sigma^{+}$ left and $\sigma^{-}$ right circular polarization at a frequency $(\omega_{\text{ref}}/2\pi)=\SI{50.1}{\kilo\hertz}$. Before entering the PEM, the linear polarization of the pump laser is adjusted to $+\SI{45}{\degree}$ to the horizontal by a high-quality Glan-Thompson polarizer. The fast axis of the PEM is oriented horizontally and the retardation is set to $\lambda/4$ for the respective pump wavelength. After passing the PEM the modulated pump beam is focused on the sample surface and locally excites electrons with a net spin polarization $S_{z}$ oriented along the sample normal $\hat{z}$.

The benefit of the PEM modulation is twofold. First, the fast, sinusoidal change of the electron spin orientation prevents the unintentional polarization of nuclear spins by transfer of angular momentum from the electron to the lattice system.\cite{OpticalOrientation} Second, the periodic modulation of the pump laser polarization enables lock-in detection of the small Kerr rotation signals.

\subsection{Lock-in Kerr rotation detection}
The spatially resolved electron spin detection by our pump-probe Kerr microscope is based on the polar MOKE: In the presence of spin-polarized electrons, the probe laser state is changed from the initially linear to elliptical polarization after reflection from the sample surface.\cite{Reim:1990} In polar geometry, the major axis of polarization is rotated by the Kerr angle $\theta$. The ratio of the major and minor axis of polarization is determined by the tangent of the Kerr ellipticity $\phi$. For the small polarization changes typically observed in pump-probe MOKE microscopy, the polar Kerr rotation depends linearly on the electron spin polarization.\cite{Kikkawa:1997} Therefore a spatially resolved measurement of the Kerr rotation $\theta(r)\propto S_{z}(r)$ can be used to map the local electron spin polarization.\cite{Crooker:2005}

The magnitude of Kerr rotations observed in pump-probe MOKE microscopy experiments is typically well below \SI{1}{\milli\radian}. To facilitate the detection of the small polarization signals we employ a balanced photoreceiver lock-in detection scheme:\cite{Crooker:1996iy,Kimel2000}

Before entering the focussing optics, the probe laser is set to vertical linear polarization by a Glan-Thompson polarizer. After reflection from the sample surface, the elliptical polarization state of the probe laser induced by the polar MOKE is described by the Jones vector:\cite{Azzam:1987}
\begin{equation}
\mathbf{E}_{\text{out}}=
\begin{pmatrix}
\text{cos}(\theta)\text{cos}(\phi)-i \text{sin}(\theta)\text{sin}(\phi)\\
\text{sin}(\theta)\text{cos}(\phi)+i \text{cos}(\theta)\text{sin}(\phi)
\end{pmatrix}
\end{equation}
After being collected by the microscope objective, the major axis of polarization of the reflected probe beam is rotated by $\SI{45}{\degree}$ by a Soleil-Babinet compensator. The compensator therefore is set to half-wave retardation and the fast axis is tilted by \SI{22.5}{\degree} to the vertical. 

We use a Wollaston prism to split the probe laser into two separate, linearly polarized beams carrying the $\hat{x}$ (horizontal) and $\hat{y}$ (vertical) polarization components which are focused on two separate photodiodes of a balanced photodetector. A high-transmission Semrock RazorEdge ultra-steep longpass filter with an appropriate cut-off wavelength is placed in front of the Wollaston prism to prevent the collinear, modulated pump laser from entering the detector. The custom-built photodetector used in our instrument is specifically optimized for raster-scanning microscopy to achieve highest possible spatial homogeneity in the detection efficiency. We therefore employ large-area Hamamatsu Photonics photodiodes to compensate for the small lateral displacement of the probe beam focus position on the diode surface caused by the the raster-scanning of the probe beam position on the sample.

The two probe beam components evoke photocurrents which are proportional to the respective beam intensity $I_{x,y}\propto\vert \left(\mathbf{E}_{\text{out}}\right)_{x,y} \vert^{2}$. A high-sensitivity transimpedance amplifier is used to generate a voltage signal which is proportional to the photocurrent difference:
\begin{equation}
V_{\text{out}}\propto \left(I_{x}-I_{y}\right)\propto \bigg(\text{sin}\left[2(\theta-\phi)\right]+\text{sin}\left[2(\theta+\phi)\right]\bigg)
\end{equation}

The periodic modulation of the optical spin excitation by the PEM leads to a sinusoidal time-dependence of the Kerr rotation and ellipticity, i.e. $\theta=\theta_{K}\text{sin}(\omega_{\text{ref}}t)$ and $\phi=\phi_{K}\text{sin}(\omega_{\text{ref}}t)$. The photodetector output voltage can be expanded in terms of odd harmonics of the PEM frequency $\omega_{\text{ref}}$ using the Jacobi-Anger identity\cite{Abramowitz:1965} as:
\begin{equation}
\label{LIinput}
\begin{aligned}
&V_{\text{out}}(t)\propto&\\ 
&\sum_{n=0}^{\infty}\bigg(J_{2n+1}\left[2(\theta_{k}-\phi_{k})\right]+J_{2n+1}\left[2(\theta_{k}+\phi_{k})\right]\bigg)\\
&\times\quad\text{sin}\left[(2n+1)\omega_{\text{ref}}t\right]
\end{aligned}
\end{equation}
where $J_{k}$ denotes the Bessel function of $k$th order.

The voltage $V_{\text{out}}(t)$ is demodulated by a Stanford Research SR530 lock-in amplifier. The lock-in output voltage $V_{\text{LI}}$ is proportional to the $\omega_{\text{ref}}$ frequency component of the input signal, i.e. the lock-in is only sensitive to the $n=0$ term of the series in Eq. \eqref{LIinput}. Furthermore, since $\vert\theta_{K}\vert\ll 1$ and $\vert\phi_{K}\vert\ll 1$, the approximation $J_{1}(x)\approx\frac{x}{2}$ can be used.\cite{Abramowitz:1965} Therefore the lock-in output voltage is a direct measure of the local Kerr rotation, i.e. $V_{\text{LI}}(r)\propto \theta_{K}(r)$. 

Finally, by repeatedly raster-scanning the probe with respect to the pump spot position and measuring the local Kerr rotation $\theta_{K}(\Delta t, r)$ as a function of the pump-probe delay $\Delta t$, full information on the spatio-temporal dynamics of the electron spin polarization $S_{z}(\Delta t, r)$ is obtained. This is the working principle of time-resolved real-space imaging of electron spins by pump-probe Kerr microscopy.

\section{Instrument characterization}
\subsection{Spatial resolution}
\begin{figure*}[tb]
\includegraphics[width=1.41\columnwidth]{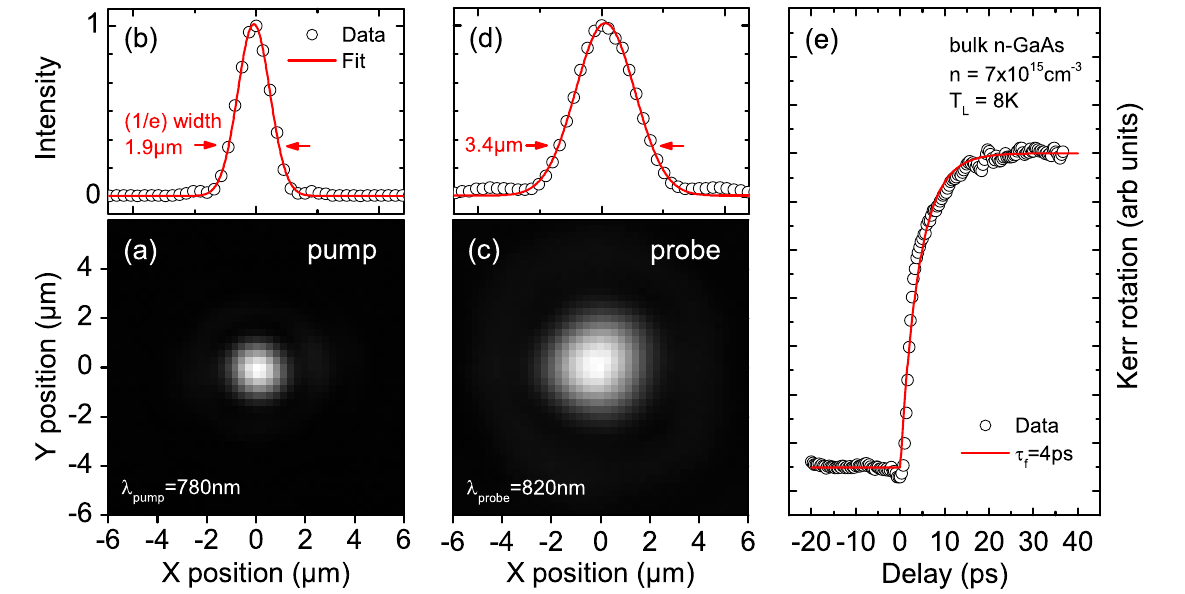}
\caption{Characterization of the spatial and temporal resolution of the pump-probe MOKE microscope. (a,c) CMOS camera images of the focused pump and probe laser spots. (b,d) Normalized intensity profiles of the pump and probe laser obtained from line cuts through the CMOS camera images. Red solid lines are Gaussian fits from which we determine the spatial resolution of the instrument. (e) Transient Kerr rotation $\theta_{K}(\Delta t)$ in a n-GaAs epilayer at $T_{L}=\SI{8}{\kelvin}$. Red solid line is a fit to an exciton formation model from which we obtain an exciton formation time  $\tau_{f}\approx\SI{4}{\ps}$ (compare text).}
\label{fig3}
\end{figure*}
We determine the spatial resolution of the instrument from CMOS camera images of the focused pump and probe spots. The square pixel size of the camera is $d_{\text{CMOS}}=\SI{5.3}{\micro\meter}$. For the $(\SI{75}{\mm}/\SI{4}{\mm})$ ratio of the focal lengths of the camera lens and the microscope objective, one pixel corresponds to a distance of \SI{0.28}{\micro\meter} on the sample surface.

From Eq. \eqref{ResolutionLimitEq} the diffraction limit for the attainable spatial resolution depends on the laser wavelength $\lambda$. In the present configuration the Kerr microscope is optimized for the investigation of GaAs-based heterostructures whose excitonic Kerr resonances are in the $\SI{810}{\nm}-\SI{820}{\nm}$ spectral range. We therefore characterize the spatial resolution for an above-bandgap central pump wavelength $\lambda_{\text{pump}}=\SI{780}{\nm}$ and a probe wavelength $\lambda_{\text{probe}}=\SI{820}{\nm}$.

In Figs. \ref{fig3} (a,c) we show CMOS camera images of the focused pump and probe spots. Line cuts through the center of the radial symmetric intensity profiles are shown in Figs. \ref{fig3} (b) for the pump and (d) for the probe spot. The intensity profiles $I(r)$ are well described by a Gaussian
\begin{equation}
I(r)=I_{0}\times\text{exp}\left(-\frac{r^{2}}{\Delta^{2}}\right)
\end{equation}
where $I_{0}$ is the maximum intensity and $\Delta$ the $(1/e)$ half-width of the spot. Gaussian fits of the spot intensity profiles shown in Figs. \ref{fig3} (b,d) yield $\Delta_{\text{Pump}}=\SI{0.95}{\micro\meter}$ and $\Delta_{\text{Probe}}=\SI{1.7}{\micro\meter}$ for the pump and probe beam. The increased probe spot width is mainly caused by the additional passage of the two aspheric lenses. We have verified that the determination of spot widths from CMOS camera images yields identical results as the conventional knife-edge scan technique. The net optical resolution $\Delta$ of the microscope is determined by the convolution of both intensity profiles. From the convolution of the Gaussian intensity profiles we obtain a spatial resolution $\Delta=\sqrt{\Delta_{\text{Probe}}^{2}+\Delta_{\text{Pump}}^{2}}=\SI{1.9}{\micro\meter}$. 

\subsection{Time resolution}
Dispersion in the tapered fiber leads to a significant prolongation of the initial $\lesssim\SI{300}{\fs}$ infrared pump pulse. As a result, the total duration of the supercontinuum pulse is of the order of \SI{5}{\ps}. However, as revealed by cross-correlation frequency-resolved optical gating (XFROG) characterization,\cite{Linden:1998} the temporal width of the arrival time distribution of individual spectral components of the supercontinuum with bandwidths comparable to our filtered pump and probe laser beams is typically of the order of \SI{500}{\fs} (Refs. \onlinecite{Dudley:2006,Metzger:2010}). From this consideration we expect an overall time resolution of  $\approx\SI{1}{\ps}$ for our pump-probe MOKE microscope.

The comparatively weak intensities of the filtered pump and probe beams impede auto-correlation measurements of the respective pulse lengths for the direct determination of the time resolution of our instrument. Furthermore, XFROG and auto-correlation measurements require additional equipment and involve changes in the experimental setup. As an alternative, we therefore determine an upper limit for the time resolution by observing the very fast initial rise of the transient Kerr rotation signal following the pulsed optical spin excitation. This approach has the advantages of being available \textit{in situ}, i.e. it does not require any modification of the experimental setup or additional equipment, and considers all optical components of the Kerr microscope in its operational configuration which could  potentially impair the time resolution of the instrument.

In Fig. \ref{fig3} (e) we show a typical time trace of the Kerr rotation $\theta_{K}(\Delta t)$ which we measure in a MBE-grown n-GaAs epilayer (room temperature electron density $n=\SI{7E15}{\per\cubic\centi\meter}$, \SI{1}{\micro\meter} layer thickness) for short delays between \SI{-20}{\ps} and \SI{+40}{\ps}. The sample temperature is $T_{L}=\SI{8}{\kelvin}$. For the detection of the Kerr rotation transient the probe laser is tuned to the excitonic resonance at $\lambda_{\text{probe}}=\SI{820}{\nm}$. The pump wavelength is $\lambda_{\text{pump}}=\SI{780}{\nm}$.

Following the excitation at $\Delta t=\SI{0}{\ps}$ we observe a very steep rise of the Kerr rotation $\theta_{K}(\Delta t)$. With increasing $\Delta t$, the increase in the Kerr rotation signal gradually slows down. We observe a $(1/e)$ rise time $t_{e}=\SI{4}{\ps}$ for which $\theta_{K}(t_{e})$ has reached \SI{63}{\percent} of the maximum amplitude. The time $t_{e}$ can be used as a rather conservative estimate for an upper limit for the time resolution of our instrument. However, we note that on the examined picosecond timescale the negative-delay flank of the initial Kerr rotation transient does not exhibit the Gaussian error function type shape which is characteristic for a time resolution limited observation of a transient optical nonlinearity. This suggests an overall time resolution $\lesssim\SI{1}{\ps}$ for our Kerr microscope. Indeed, in the present measurement, the instruments time resolution is partially masked by the finite exciton formation time which limits the Kerr rotation rise time:

The gradual slowing of the rise of $\theta_{K}(\Delta t)$ observed for delays $\Delta t\gtrsim\SI{4}{ps}$ likely stems from the non-instantaneous exciton formation following the non-resonant optical electron spin excitation. In bulk GaAs, the photo-induced Kerr rotation results from an energetic splitting $\Delta E$ of the $E_{0}$ excitonic resonances for $\sigma^{\pm}$ light, the excitonic spin splitting.\cite{Vina:1996,Kozhemyakina:2009,Henn:2013} The size of the spin splitting determines the magnitude of the Kerr rotation, i.e. $\theta_{K}\propto\Delta E$. Since $\Delta E$ is proportional to the exciton density $n_{X}$,\cite{Vina:1996} we can obtain an estimate of the exciton formation time $t_{f}$ from the Kerr rotation transient $\theta_{K}(\Delta t)$: Following Refs. \onlinecite{Kozhemyakina:2009,Kozhemyakina:2010} we model the exciton formation for very short delays $\Delta t>0$ by a time-dependent rise of the exciton density $n_{X}(\Delta t)\approx n_{X,\text{max}}[1-\text{exp}(-\Delta t/t_{f})]$. From the fit of $\theta_{K}(\Delta t) \propto n_{X}(\Delta t)$ shown in Fig. \ref{fig3} (e) we determine an exciton formation time $t_{f}=\SI{4.0(3)}{\ps}$ under our non-resonant excitation conditions. This finite formation time is causing the comparatively slow rise in the Kerr rotation transient which masks the instruments faster time resolution.

\section{Measurements and discussion}
\label{measurement}
To demonstrate the operation of our Kerr microscope we present time-resolved real-space imaging measurements of the diffusion of optically excited electron spin packets in a (110) GaAs QW. The exploration of the strongly anisotropic electron spin relaxation rates in (110) QWs has received considerable attention in the past.\cite{Ohno:1999,Dohrmann:2004,Belkov:2008,Muller:2008,Griesbeck:2012} In contrast, only a small amount of work has been devoted to the experimental investigation of electron spin transport in such (110) QWs. Electron spin diffusion coefficients in (110) QWs have mainly been determined by transient spin grating measurements\cite{Eldridge:2008,Hu:2011,Chen:2012} which have focused on elevated temperatures. As a first step towards closing this gap we present low-temperature electron spin diffusion coefficients for a (110) GaAs QW sample which we determine directly from time-resolved pump-probe MOKE microscopy.

The sample has been grown by molecular beam epitaxy (MBE) on a semi-insulating (110)-oriented GaAs substrate. Growth was initiated by a \SI{200}{\nm} GaAs buffer layer followed by \SI{500}{\nm} of $\text{Al}_{0.15}\text{Ga}_{0.85}\text{As}$ and a series of five identical \SI{20}{nm} wide GaAs QWs which are embedded in \SI{64}{\nm} $\text{Al}_{0.15}\text{Ga}_{0.85}\text{As}$ barriers. The sample is capped by a \SI{120}{\angstrom} $\text{Al}_{0.15}\text{Ga}_{0.85}\text{As}$ layer followed by a top \SI{2}{\nm} GaAs layer. Our standard continuous-wave Hanle-MOKE\cite{Volkl:2011} characterization and resonant spin amplification\cite{Glazov:2008,Griesbeck:2012} (RSA) measurements (not shown here) have revealed long out-of-plane spin lifetimes $T_{s}^{z}\gtrsim\SI{30}{\ns}$ at low $T_{L}$ which suggests the presence of a low-density two-dimensional electron gas (2DEG) in the QWs.

For all measurements presented here, the central probe laser wavelength is tuned to the (1e-1hh) excitonic resonance at $\lambda_{\text{probe}}=\SI{814}{\nm}$ and the time-averaged probe laser power is $P_{\text{probe}}=\SI{12}{\micro\watt}$. The pump central laser wavelength is $\lambda_{\text{pump}}=\SI{780}{\nm}$ and the pump power is varied between $P_{\text{pump}}=\SI{1.5}{\micro\watt}$ and $\SI{5}{\micro\watt}$ as indicated below. 

At low $T_{L}$, the spin lifetime $T_{s}^{z}$ is comparable to our pulse-to-pulse interval $t_{\text{rep}}=\SI{27.5}{\ns}$. At short negative delays, i.e. before the arrival of the respective pump pulse, we therefore observe a small remanent spin polarization which results from excitation by previous pump pulses. To avoid repetition-rate artifacts in the following analysis we apply a small in-plane magnetic field $B_{xy}\approx\SI{2}{\milli\tesla}$ which suppresses previous-pulse contributions by destructive RSA\cite{Kikkawa:1998}.

Following pulsed local excitation by a Gaussian pump spot, the time evolution of the lateral expansion of the electron spin packet in the (110) QW plane is:\cite{Smith:1988,Yoon:1992,Zhao:2009}
\begin{equation}
\label{DiffusionSol}
S_{z}(\Delta t,r) = \frac{\sigma_{0}^{2}S_{z\text{,}0}}{\sigma_{0}^{2}+4D_{s}\Delta t}\text{exp}\left(-\frac{r^{2}}{w_{s}^{2}}-\frac{\Delta t}{T^{z}_{s}}\right)
\end{equation}
with an initial $(1/e)$ half width $\sigma_{0}$ and amplitude $S_{z\text{,}0}$. 

While spin relaxation leads to an decrease of the overall amplitude of the Gaussian spin packet, the squared $(1/e)$ half-width
\begin{equation}
\label{timeDepWidth}
w_{s}^{2}=\sigma_{0}^{2}+4D_{s}\Delta t
\end{equation}
only depends on the spin diffusivity $D_{s}(\Delta t)$ and increases linearly with time. To determine the electron spin diffusion coefficient of the (110) QWs we measure a series of spin polarization profiles $S_{z}(\Delta t,r)$ for increasing delays $\Delta t$ and extract the spin packet width $w_{s}(\Delta t)$. From Eq. \eqref{timeDepWidth} we then directly obtain the spin diffusivity from the slope of a linear fit of the time-dependence of the spin packet width. 

\begin{figure}[tb]
\includegraphics[width=1\columnwidth]{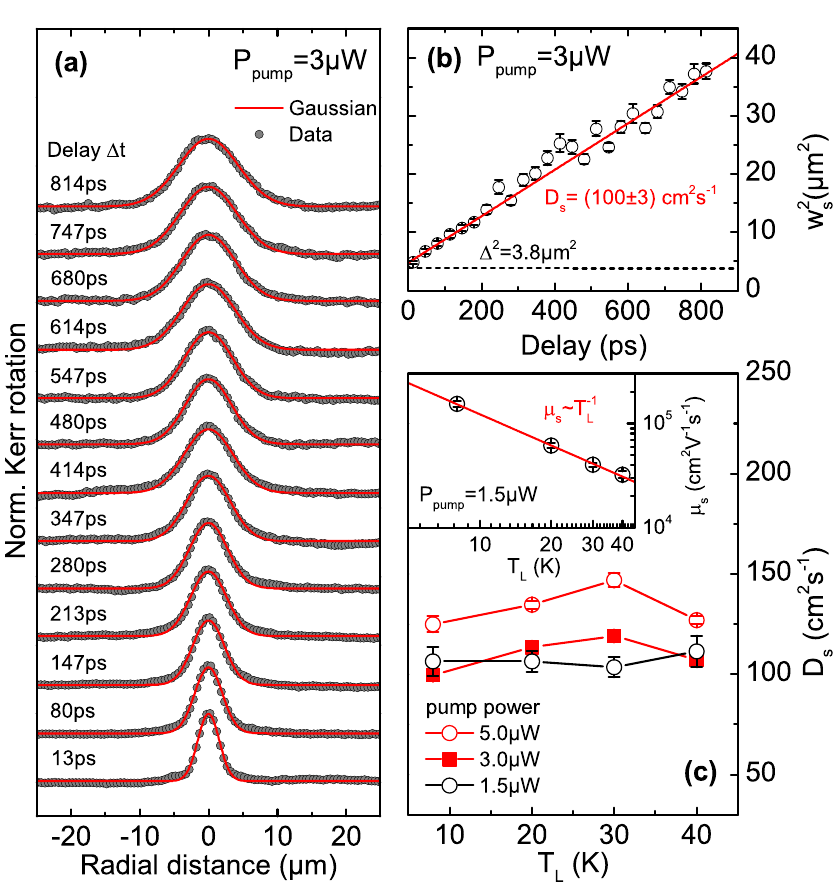}
\caption{(a) Time-evolution of an optically excited electron spin packet in the (110) QW sample at $T_{L}=\SI{8}{\kelvin}$. The diffusive spread of the spin packet is observed from the increase in the profile width for increasing delays (bottom to top; the delay $\Delta t$ is indicated above the respective curve). Markers are experimental data, red solid lines are Gaussian fits. (b) Linear increase of the squared (1/e) half width $w_{s}^{2}(\Delta t)$. For $\Delta t=\SI{0}{\ps}$ the spin packet width coincides with the optical resolution $\Delta^2=\SI{3.8}{\micro\meter\squared}$ indicated by the black dashed line. The spin diffusion coefficient $D_{s}=\SI{100(3)}{\centi\meter\squared\per\second}$ is obtained from the slope of a linear fit (red solid line). (c) Lattice temperature dependence of $D_{s}$ for different pump powers. Inset: Spin mobility $\mu_{s}(T_{L})$ calculated via the Einstein relation from $D_{s}(T_{L})$ measured for the lowest pump power.}
\label{fig4}
\end{figure}
In Fig. \ref{fig4} (a) we exemplarily show electron spin polarization profiles $S_{z}(\Delta t, r)$ which we measure at $T_{L}=\SI{8}{\kelvin}$ with a pump power $P_{\text{pump}}=\SI{3}{\micro\watt}$ for delays up to \SI{814}{\ps} together with Gaussian fits. The shape of the spin packet remains Gaussian for all times and the diffusive spread of the spin packet with increasing $\Delta t$ is clearly observed. The linear increase of the squared spin packet width $w_{s}^{2}(\Delta t)$ expected from Eq. \eqref{timeDepWidth} is confirmed in Fig. \ref{fig4} (b). From the slope of a linear fit (red solid line) we obtain a spin diffusion coefficient $D_{s}=\SI{100(3)}{\centi\meter\squared\per\second}$.

Following the above procedure, we measure the spin diffusivity $D_{s}$ for $T_{L}$ between \SI{8}{\kelvin} and \SI{40}{\kelvin} and pump powers $P_{\text{pump}}=\SI{1.5}{\micro\watt}$, \SI{3}{\micro\watt}, and \SI{5}{\micro\watt}. Our results for the dependence of $D_{s}$ on excitation density and lattice temperature are summarized in Fig. \ref{fig4} (c). 

For all examined pump powers, we find that $D_{s}$ only varies weakly with $T_{L}$. We observe that reducing $P_{\text{pump}}$ from \SI{5}{\micro\watt} to \SI{3}{\micro\watt} results in a systematic decrease of the spin diffusion coefficient.  However, an additional reduction to a very low pump power of \SI{1.5}{\micro\watt} does not result in a further significant decrease of $D_{s}$. This suggests that we have reduced  pump-induced disturbances sufficiently to observe the intrinsic temperature dependence of the spin diffusivity. We therefore focus on $D_{s}(T_{L})$ obtained for the lowest pump power of \SI{1.5}{\micro\watt} for the following discussion.

Contrary to the case of bulk GaAs,\cite{Quast:2013} we observe a constant diffusivity $D_{s}\approx\SI{100}{\centi\meter\squared\per\second}$ which does not increase with $T_{L}$. A similar temperature-independent spin diffusion coefficient has also been observed previously by transient spin-grating measurements in a (110) GaAs QW for elevated $T_{L}$ between \SI{60}{\kelvin} and room temperature.\cite{Eldridge:2008} Following Ref. \onlinecite{Eldridge:2008} we use the Einstein relation to calculate the spin mobility $\mu_{s}=(e/k_{B}T_{L})D_{s}$ of the non-degenerate (110) QW 2DEG. We then use the temperature dependence $\mu_{s}(T_{L})$ to infer the dominant scattering mechanism which limits the spin diffusivity.

From the observed constant diffusivity $D_{s}$ and the Einstein relation, we expect a spin mobility $\mu_{s}\propto T_{L}^{-1}$, which we indeed observe, as shown in the inset of Fig. \ref{fig4} (c). This temperature dependence of $\mu_{s}$ is characteristic for a non-degenerate 2DEG for which the mobility is limited by scattering with acoustic phonons for which the probability increases $\propto (k_{B}T_{L})$ (Ref. \onlinecite{Eldridge:2008}).

\section{Summary}
We have presented a supercontinuum fiber-laser pump-probe scanning MOKE microscope which enables the investigation of spin relaxation and spin transport phenomena in semiconductor heterostructures with picosecond time and micrometer spatial resolution at cryogenic temperatures. The pulsed supercontinuum laser source provides access to the whole visible and near-infrared spectral range, therefore offering the possibility to investigate a wide range of different semiconductor material systems. By implementing two separate spectral filtering schemes we have obtained the ability to independently select the pump and continuously tunable probe laser wavelength. 

We have demonstrated the capabilities of the Kerr microscope by providing an investigation of low-temperature electron spin propagation in (110) GaAs QWs. We have measured electron spin diffusion coefficients of a low-density (110) QW 2DEG by time-resolved real-space imaging of the time-evolution of an optically excited electron spin packet. From the temperature dependence of the spin mobility we have identified acoustic phonon scattering as the dominant scattering mechanism which limits the low-temperature electron spin transport in the (110) QWs.

We anticipate a broad variety of future applications for the MOKE microscope presented here. The high spatial and temporal resolution of our instrument allows for the investigation of spin transport in both standard semiconductor heterostructures and lithographically defined spin-transport devices. The ability to independently change the pump and probe energy further enables us to control the electron temperature by variation of the excitation excess energy. Our microscope thereby offers the possibility to study the interplay of charge, spin and heat transport in semiconductors which is the subject of the emerging field of \textit{spin caloritronics}.\cite{Bauer:2010,Bauer:2012}

\begin{acknowledgments}
The authors gratefully acknowledge support by the DFG (SPP1285 OS98/9-3 and SPP1285 SA598/8-3).
\end{acknowledgments}
%
\end{document}